\pdfoutput=1 

\documentclass{mcom-l}

\usepackage{siunitx}
\usepackage{fancyhdr,graphicx,amsmath,amssymb}
\usepackage[ruled,vlined]{algorithm2e}
\include{pythonlisting}
\usepackage{booktabs}
\usepackage[table,xcdraw]{xcolor}
\usepackage{subfiles}
\usepackage{fixltx2e}
\usepackage{setspace}
\usepackage{soul}
\usepackage{mathtools}
\DeclarePairedDelimiter{\ceil}{\lceil}{\rceil}
\usepackage{tikz}
\usetikzlibrary{shapes.geometric,calc}
\usepackage{multirow}
\usepackage{comment}
\usepackage{url}
\usepackage{hyperref}
\usepackage[backend=bibtex,style=numeric]{biblatex} 
\theoremstyle{definition}

\theoremstyle{remark}

\numberwithin{equation}{section}

\begin{document}

\title[Variable selection with multiply-imputed datasets]{Variable selection with multiply-imputed datasets: choosing between stacked and grouped methods}


\author[Du et al.]{Jiacong Du$^{1}$}

\thanks{$^{1}$ Department of Biostatistics, University of Michigan, Ann Arbor, Michigan}

\author[]{Jonathan Boss$^{1}$}

\author[]{Peisong Han$^{1}$}

\author[]{Lauren J Beesley$^{1}$}

\author[]{Stephen A Goutman$^{2}$}
\thanks{$^{2}$ Department of Neurology, University of Michigan, Ann Arbor, Michigan}

\author[]{Stuart Batterman$^{3}$}
\thanks{$^{3}$ Department of Environmental Health Science, University of Michigan, Ann Arbor, Michigan}

\author[]{Eva L Feldman$^{2}$}

\author[]{Bhramar Mukherjee$^{1}$}


\date{}

\dedicatory{}

\begin{abstract}
Penalized regression methods, such as lasso and elastic net, are used in many biomedical applications when simultaneous regression coefficient estimation and variable selection is desired. However, missing data complicates the implementation of these methods, particularly when missingness is handled using multiple imputation. Applying a variable selection algorithm on each imputed dataset will likely lead to different sets of selected predictors, making it difficult to ascertain a final active set without resorting to ad hoc combination rules.\\ 
\indent In this paper we consider a general class of penalized objective functions which, by construction, force selection of the same variables across multiply-imputed datasets. By pooling objective functions across imputations, optimization is then performed jointly over all imputed datasets rather than separately for each dataset. We consider two objective function formulations that exist in the literature, which we will refer to as "stacked" and "grouped" objective functions. Building on existing work, we (a) derive and implement efficient cyclic coordinate descent and majorization-minimization optimization algorithms for both continuous and binary outcome data, (b) incorporate adaptive shrinkage penalties, (c) compare these methods through simulation, and (d) develop an R package \textit{miselect} for easy implementation. Simulations demonstrate that the "stacked" objective function approaches tend to be more computationally efficient and have better estimation and selection properties. We apply these methods to data from the University of Michigan ALS Patients Repository (UMAPR) which aims to identify the association between persistent organic pollutants and ALS risk.


\end{abstract}

\maketitle

\section{Introduction}

Variable selection in the presence of missing data is a frequent issue in statistical analysis, however, there is a surprising lack of methods and tools to address this problem in a pragmatic and principled way. The most common approaches in the literature can be characterized as "quick fixes", namely complete-case analysis, single imputation, and ad hoc rules for harmonizing selection across multiply-imputed datasets \cite{wood2008, lachenbruch2011, ghosh2015}. \citeauthor*{wood2008} \cite{wood2008} outlined three natural ad hoc rules for finalizing selection after identifying the active set of predictors in each imputed dataset. These rules consider a variable to be selected: if it is selected in at least one imputed dataset, if it is selected in all imputed datasets, and if it is selected in at least half of the imputed datasets. Other existing approaches can be divided into the following categories: Bayesian methods utilizing the data augmentation strategy of Tanner and Wong (1987) \cite{tanner1987calculation, yang2005imputation, ibrahim2008bayescox}, pooled posterior inclusion probabilities \cite{yang2005imputation}, bootstrapped frequentist inclusion probabilities \cite{heymans2007variable, long2015, liu2016}, variable selection on the stacked imputed datasets \cite{wood2008, wan2015variable}, group lasso regularized pooled objective functions \cite{chen2013variable, geronimi2017variable, marino2017}, inverse probability weighting (IPW)  \cite{johnson2018}, low-rank matrix completion \cite{candes2010power, choi2013investigation}, and lasso regularized inverse covariance estimation \cite{loh2012, stadler2012}.  Expectation-Maximization (EM) algorithms for penalized regression and new information criteria for model selection defined by the EM principle of integrating the augmented likelihood over the missing data distribution have also been proposed \cite{claeskens2008variable, ibrahim2008model, garcia2010general, garcia2010variable, shen2012model, sabbe2013}.

A problem with the Bayesian methods, the EM-algorithm approaches, and the EM-style information criteria is that a custom Markov chain Monte Carlo sampler or optimization algorithm is required for each analysis. Pooled posterior inclusion probabilities and bootstrapped frequentist inclusion probabilities are targeted towards selection but are not designed to generate point estimates of the selected regression coefficients. The IPW approach of \cite{johnson2018} assumes a monotone missingness structure and inherits the known stability issues surrounding IPW-based procedures \cite{johnson2018}. Low-rank matrix completion is an interesting deterministic approach, but implicit assumptions on the missing data structure are unclear. Additionally, low-rank matrix completion assumes that there exists a good low-rank representation of the design matrix, which may not always be the case \cite{choi2013investigation}. Lasso-regularized inverse covariance estimation, as outlined in \cite{loh2012}, assumes that the missing data mechanism is missing completely at random and that the probability of missingness is a fixed constant across all covariates and observations. These are strong restrictions within the scope of missing data problems \cite{loh2012, little2019statistical}. Alternatively, lasso-regularized inverse covariance estimation presented in \citeauthor{stadler2012} directly uses the observed-data log-likelihood under the assumption that the outcome and the covariates jointly follow a multivariate normal distribution \cite{stadler2012}. Often, the observed data likelihood is analytically intractable, particularly for non-normally distributed outcomes. 

The  imputation stacking and the grouped objective function approaches are appealing because they handle variable selection and estimation given previously imputed datasets obtained from standard or custom multiple imputation software. Such data are often released by national entities like the National Health and Nutrition Examination Survey \cite{iranpour2019association}. Additionally, no ad hoc pooling is required to determine the final active set. Both approaches allow simultaneous model selection and estimation and can generate interpretable regression coefficient estimates. The stacking approach can be reformulated in terms of maximizing an objective function pooled across imputed datasets, where parameter/regression coefficient values are assumed to be equal across imputed datasets \cite{wan2015variable}. The grouped objective function approach can then be viewed as a generalization of the imputation stacking strategy, where parameter values are allowed to differ across imputed datasets but selection will be consistent across imputed datasets by incorporating a grouped penalty. We will refer to these two approaches of stacking and grouping as the homogeneous and heterogeneous objective function strategies, respectively. 

Despite the many advantages of these selection and estimation strategies, implementing selection using pooled objective function methods has been limited in practice. Existing approaches are mostly targeted toward normally-distributed outcomes and very few approaches have been extended to discrete outcomes. The algorithm in \citeauthor{chen2013variable} \cite{chen2013variable} for optimizing heterogeneous pooled objective functions involves a local quadratic approximation, which converts the heterogeneous pooled objective function into multiple ridge regressions. This effectively forces users to set threshold in an ad hoc manner. In addition, there are certain limitations of using off-the-shelf R packages for variable selection like \textit{glmnet} \cite{hastie2014glmnet} and \textit{gcdnet} \cite{yang2017package}, even for the grouped or stacked methods. Though the \textit{gcdnet} package allows adaptive penalties for each coefficient, it lacks the flexibility to treat observations with different weights. Additionally, little work has been done to compare the performance of stacked and grouped methods.

In this paper, we extend existing work on pooled objective functions for handling variable selection with missing data to incorporate binary outcomes and adaptive penalties (Section 2). The extension to binary outcome data is crucial given that many health outcomes of interest are binary, e.g., diseased versus non-diseased. We derive novel cyclic coordinate descent algorithms for optimizing homogeneous pooled objective functions and majorization-minimization (MM) algorithms coupled with block coordinate descent updates to obtain optimizers of heterogeneous pooled objective functions for both continuous and binary outcome data (Section 3). Unlike existing algorithms, our proposed methods provide exact shrinkage to zero without any ad hoc thresholding. We provide an R package \textit{miselect}, available from: \url{https://github.com/umich-cphds/miselect/releases/tag/v0.7}, allowing users to easily implement the proposed methods. In our motivating example, we apply these methods to identify persistent organic pollutants (POPs) associated with Amyotrophic Lateral Sclerosis (ALS) susceptibility using data collected from the University of Michigan ALS Patients Repository (Section 4). Seventeen of the 23 POPs have between 10\% to 60\% below their respective detection limits, making complete-case analysis infeasible. Finally, through a simulation study, we compare the performance of the proposed methods in terms of variable selection and estimation accuracy (Section 5).

\section{Methods}
 \label{section:methods}

Let $\boldsymbol{X}_d$ denote the $n \times p$ matrix of predictor variables and $\boldsymbol{Y}_d$ be the $n \times 1$ vector of responses for the $d$-th imputed dataset ($d = 1, \hdots D$). Let $\boldsymbol{X}_{d,i}$ indicate the $p \times 1$ covariate vector for the $i$-th observation in the $d$-th imputed dataset, $Y_{d,i}$ be the response for the $i$-th observation in the $d$-th imputed dataset, and $X_{d,ij}$ denote the $j$-th covariate for the $i$-th observation in the $d$-th imputed dataset. The $p \times 1$ vector of regression coefficients for the $d$-th dataset is given by $\boldsymbol{\beta}_d$, the regression coefficient in the $d$-th dataset corresponding to the $j$-th covariate is denoted by $\beta_{d,j}$, and the intercept parameter for the $d$-th imputed dataset is $\mu_d$. The vector of regression parameters for the $d$-th dataset is given by $\boldsymbol{\theta}_d = (\mu_d, \boldsymbol{\beta}_d)$. Under this general framework, we present various approaches to identify active set of predictors.

\subsection{Penalized regression on individual datasets}

A common approach in practice is to use ad hoc rules for determining the final set of selected variables. Often this proceeds by fitting a penalized regression procedure on each imputed dataset separately:

\begin{equation}\label{eq:1}
    \hat{\boldsymbol{\theta}}_d = \operatorname*{arg min}_{\boldsymbol{\theta}_d} \ \bigg\{-\frac{1}{n}\sum_{i=1}^{n}logL(\boldsymbol{\theta}_d|Y_{d,i},\boldsymbol{X}_{d,i}) + \lambda P_{\alpha}(\boldsymbol{\beta}_d) \bigg\}
\end{equation}

\noindent for $d = 1,...,D$ where $logL$ is the log-likelihood function, $P_\alpha(\boldsymbol{\beta}_d)$ is the penalty function of $\boldsymbol{\beta}_d$ parameterized by $\alpha$ and $\lambda \in [0,\infty)$ is a tuning parameter that controls the relative contribution of the penalty.

In this paper, we will focus on four penalty functions: LASSO \cite{tibshirani1996regression}, adaptive LASSO (aLASSO) \cite{zou2006adaptive}, elastic net (ENET) \cite{zou2005regularization} and adaptive elastic net (aENET) \cite{zou2009adaptive}. The penalty functions can be expressed as:

\begin{enumerate}
    \item LASSO: $ P_{\alpha}(\boldsymbol{\beta}_d) =  \sum_{j=1}^{p} |\beta_{d,j}|$
        
    \item aLASSO : $ P_{\alpha}(\boldsymbol{\beta}_d) =  \sum_{j=1}^{p} \hat{a}_{d,j} |\beta_{d,j}|$
        
    \item ENET: $ P_{\alpha}(\boldsymbol{\beta}_d) =  \alpha \sum_{j=1}^{p} |\beta_{d,j}| +  (1-\alpha)\sum_{j=1}^{p} \beta_{d,j}^2$
        
    \item aENET: $ P_{\alpha}(\boldsymbol{\beta}_d) =  \alpha \sum_{j=1}^{p} \hat{a}_{d,j} |\beta_{d,j}| +  (1-\alpha)\sum_{j=1}^{p} \beta_{d,j}^2$ 
\end{enumerate}

\noindent The intuitive appeal of adaptive weights is that they allow differential penalization of each covariate based on an initial estimate of the regression coefficient vector (smaller $\hat{\beta}_{d,j}^{0}$ implies a harsher penalty on $\beta_{d,j}$). Moreover, adaptive penalties are known to address estimation and selection consistency issues that have been observed for non-adaptive penalties \cite{zou2006adaptive, zou2009adaptive}. Here, the adaptive weight $\hat{a}_{d,j} = \big( |\hat{\beta}_{d,j}^{0}| + 1/n \big) ^{-\gamma}$ for some $\gamma >0$, where $\hat{\beta}_{d,j}^{0}$ is an initial estimate of $\beta_{d,j}$, often determined from ordinary least squares (OLS) or maximum likelihood estimation when $p$ is smaller than $n$. If $p$ is larger than $n$, $\hat{\beta}_{d,j}^{0}$ can be obtained using LASSO or ENET depending on the correlation structure of the predictors. For the purposes of this paper, we use ENET initial values to calculate the adaptive weights for both aLASSO and aENET. To avoid tuning on $\gamma$, we follow \cite{zou2009adaptive}, which fixes $\gamma = \ceil{2v/1-v}+1$ where $v = log(p)/log(n)$. The $1/n$ term prevents division by zero for initial regression coefficient estimates that are exactly equal to zero.

\begin{figure}[h]
    \centering
    \includegraphics[width=0.6\textwidth]{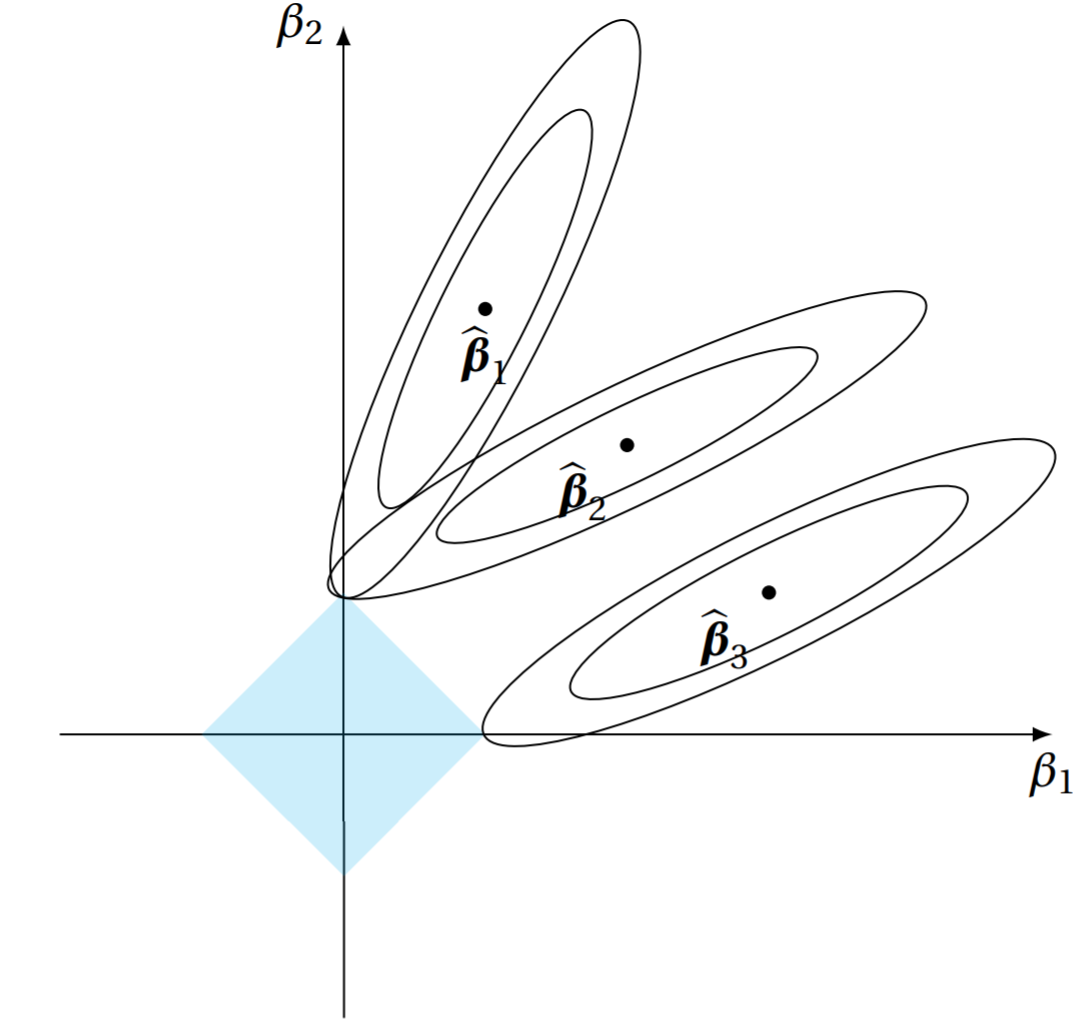}
    \caption{Example of LASSO fit separately on each imputed dataset with two predictor variables. $\widehat{\boldsymbol{\beta}}_{d}$, for $d=1,2,3$ stands for the coefficients estimates in the $d$-th imputed dataset. Each black dot represents the OLS estimates for each imputed dataset. The contours represent the data likelihood.}
    \label{lasso_individual_datasets}
\end{figure}

For illustrative purposes, suppose that there are three imputed datasets and that LASSO is fit separately on each imputed dataset (with a common $\lambda > 0$). Furthermore, assume that $\boldsymbol{X}_d$ is an $n \times 2$ matrix for all $d = 1,2,3$, i.e., we are applying LASSO shrinkage to coefficients of only two covariates. Figure \ref{lasso_individual_datasets} visualizes the geometry of the constrained region in this hypothetical scenario. Note that the OLS estimates for each of the imputed datasets are slightly different and, when shrunk to the constrained region, lead to two cases: (i) $\hat\beta_{d,1} = 0$ and $\hat\beta_{d,2} \neq 0$ and (ii) $\hat\beta_{d,1} \neq 0$ and $\hat\beta_{d,2} = 0$. This toy example illustrates the fundamental problem of trying to harmonize variable selection across imputed datasets without borrowing information across imputed datasets.

\subsection{Homogeneous pooled objective functions (stacking)}

In a homogeneous pooled objective function, we sum the objective functions for each of the imputed datasets together and jointly optimize the collective objective function:
\begin{equation}\label{eq:homogenous_base}
    \hat{\boldsymbol{\theta}} = \operatorname*{arg min}_{\boldsymbol{\theta}} \ \Bigg\{-\frac{1}{n}\sum_{d=1}^{D}\sum_{i=1}^{n}logL(\boldsymbol{\theta}|Y_{d,i},\boldsymbol{X}_{d,i}) + \lambda P_{\alpha}(\boldsymbol{\beta}) \Bigg\}
\end{equation}
\noindent Note that $\boldsymbol{\theta}$ is not indexed by $d$. This implies that optimizing the pooled objective function will result in one estimated parameter vector $\hat{\boldsymbol{\theta}}$, thereby enforcing uniform selection across all imputed datasets.

A nice feature of the homogeneous pooled objective function is that optimization is straightforward. Namely, optimization of the homogeneous pooled objective function is equivalent to stacking the imputed datasets  and fitting the desired penalized regression algorithm on the stacked imputed datasets with existing software. Therefore, the homogeneous pooled objective function provides a framework for pooling penalized regression estimates across imputed datasets for a general class of objective functions. However, stacking all imputed datasets can be viewed as artificially increasing the sample size. A common way to address this is to add an observation weight, $o_i = 1/D$, so that the total weight for each subject in the stacked dataset sums up to one. An alternative observation weight specification, proposed in \citeauthor*{wan2015variable} \cite{wan2015variable}, is $o_i = f_i/D$, where $f_i$ is the number of observed predictors out of the total number of predictors for subject $i$, which accounts for varying degrees of missing information for each subject. That being said, upweighting subjects with less missingness and downweighting subjects with more missingness can, in some sense, be viewed as making the optimization more like complete-case analysis, which might be problematic for Missing at Random (MAR) and Missing not at Random (MNAR) scenarios. Going forward, we consider both equal weights ($o_i = 1/D$) and the observation weights proposed in \citeauthor{wan2015variable} ($o_i = f_i/D$). The weighted homogeneous pooled objective function can be written as:
\begin{equation}\label{eq:homogenous_modif}
    \hat{\boldsymbol{\theta}} = \operatorname*{arg min}_{\boldsymbol{\theta}} \ \Bigg\{-\frac{1}{n}\sum_{d=1}^{D}\sum_{i=1}^{n}o_i logL(\boldsymbol{\theta}|Y_{d,i},\boldsymbol{X}_{d,i}) + \lambda P_{\boldsymbol{\alpha}}(\boldsymbol{\beta}) \Bigg\}
\end{equation}

From this stacking approach, we extend the penalty functions to LASSO, aLASSO, ENET and aENET. When the equal weight is used, the methods are named as follows with the corresponding penalty functions:

\begin{enumerate}
    \item Stacking LASSO (SLASSO):
    $ P_{\alpha}(\boldsymbol{\beta}) =  \sum_{j=1}^{p} |\beta_{j}|$
    
    \item Stacking adaptive LASSO (SaLASSO):
    $ P_{\alpha}(\boldsymbol{\beta}) =  \sum_{j=1}^{p} \hat{a}_j |\beta_{j}|$
    
    \item Stacking elastic net (SENET):
    $ P_{\alpha}(\boldsymbol{\beta}) =  \alpha \sum_{j=1}^{p} |\beta_{j}| +  (1-\alpha)\sum_{j=1}^{p} \beta_{j}^2$
    
    \item Stacking adaptive elastic net (SaENET):
    $ P_{\alpha}(\boldsymbol{\beta}) = \alpha \sum_{j=1}^{p} \hat{a}_j |\beta_{j}| +  (1-\alpha)\sum_{j=1}^{p} \beta_{j}^2$
    
\end{enumerate}
\noindent where $\hat{a}_j = \big( |\hat{\beta}_j| + 1/(nD) \big) ^{-\gamma}$ and $\hat{\beta}_j$ is estimated through SENET. Following \citeauthor*{zou2009adaptive} \cite{zou2009adaptive}, $\gamma$ is fixed to be $\ceil{2v/1-v}+1$, where $v=log(p)/log(nD)$. 

When $o_i = f_i/D$, the penalized methods are named SLASSO(w), SaLASSO(w), SENET(w) and SaENET(w), respectively. The initialization of the adaptive weights for the stacking approach with the observation weights is calculated through SENET(w).

\subsection{Heterogeneous pooled objective functions (grouping)}

An alternative to the homogeneous pooled objective function is the heterogeneous pooled objective function. This method imposes uniform variable selection across imputed datasets by adding an additional group Lasso penalty to the objective function \cite{chen2013variable, geronimi2017variable, yuan2006model}. The optimizer of the heterogeneous pooled objective function can be mathematically expressed as:
\begin{equation}\label{eq:heterogeneous_modif}
    \big(\hat{\boldsymbol{\theta}}_1,...,\hat{\boldsymbol{\theta}}_D\big) = \operatorname*{arg min}_{\boldsymbol{\theta}_1,...,\boldsymbol{\theta}_D} \ \Bigg\{-\frac{1}{n}\sum_{d=1}^{D}\sum_{i=1}^{n}logL(\boldsymbol{\theta}_d|Y_{d,i},\boldsymbol{X}_{d,i}) + \lambda P(\boldsymbol{\beta}_1, \boldsymbol{\beta}_2,...,\boldsymbol{\beta}_D) \Bigg\}
\end{equation}

\noindent where $\lambda \in [0,\infty)$ is a tuning parameter. Chen et al. (2013) originally formulated a special case of the heterogeneous pooled objective function known as MI-LASSO, where the penalty function is:

$$ P(\boldsymbol{\beta}_1, \boldsymbol{\beta}_2,...,\boldsymbol{\beta}_D) =  \sum_{j=1}^{p}\sqrt{\sum_{d=1}^{D}\beta_{d,j}^2}$$

We refer to the objective function as heterogeneous because the parameter vector is now indexed by $d$, meaning that $\boldsymbol{\theta}_1 \neq ... \neq \boldsymbol{\theta}_D$. Although the $\boldsymbol{\theta}_d$'s are not identical, for any fixed $j$, the group Lasso penalty jointly shrink all $\beta_{d,j}$'s to zero, i.e. $\beta_{1,j} = ... = \beta_{D,j} = 0$. This allows for uniform selection across imputed datasets but also allows for variability in the non-zero estimated coefficients across imputed datasets. Based on Chen et al. (2013), we consider the following penalties:

\begin{enumerate}
    \item Group LASSO (GLASSO):
    $  P(\boldsymbol{\beta}_1, \boldsymbol{\beta}_2,...,\boldsymbol{\beta}_D) =   \sum_{j=1}^{p}\sqrt{\sum_{d=1}^{D}\beta_{d,j}^2}$
    
    \item Group adaptive LASSO (GaLASSO):
    $ P(\boldsymbol{\beta}_1, \boldsymbol{\beta}_2,...,\boldsymbol{\beta}_D) =   \sum_{j=1}^{p}\hat{a}_j\sqrt{\sum_{d=1}^{D}\beta_{d,j}^2}$
    
\end{enumerate}

\noindent where $\hat{a}_j = \big(\sqrt{\sum_{d=1}^{D} \hat{\beta}_{d,j}^2} + 1/(nD) \big)^{-\gamma}$. $\hat{\beta}_{d,j}$ is estimated from GLASSO, $\gamma = \ceil{2v/1-v}+1$, and $v=log(pD)/log(nD)$ \cite{zou2009adaptive}.

{\bf Remark:} We do not extend the grouped LASSO based approaches to grouped ENET based approaches because grouping itself provides constraints similar to the ENET penalty. How to effectively account for correlations among predictors using the grouped penalty remains an open question.

\section{Optimization}

\begin{figure}[h]
    \centering
    \includegraphics[width=13cm, height=9cm]{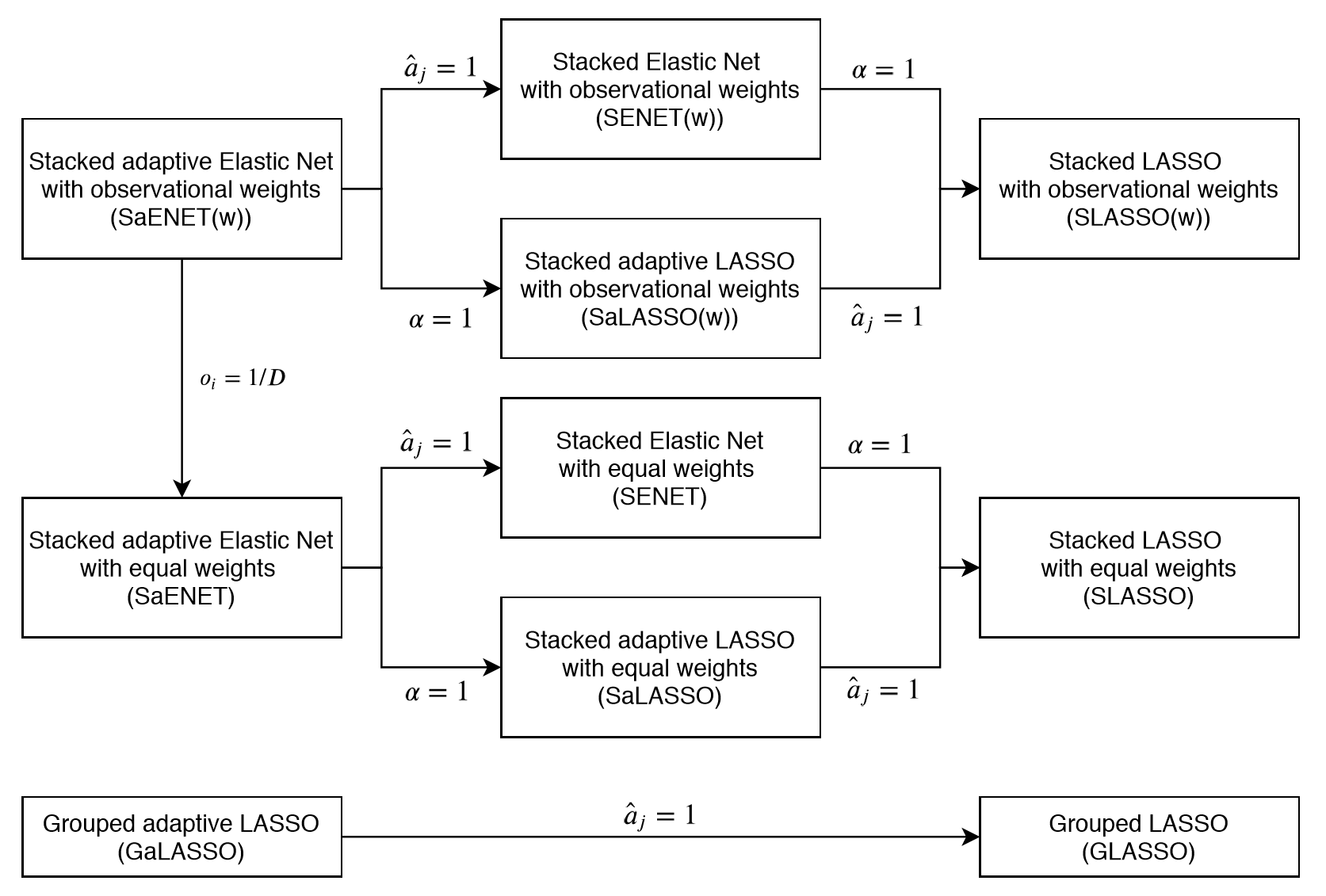}
    \caption{Illustration of how to obtain penalized pooled objective function methods, which are special cases of SaENET(w) and GaLASSO.}
    \label{flowchart}
\end{figure}

In this section, we outline the optimization routines for SaENET(w) and GaLASSO with binary outcomes. To optimize the SaENET(w) objective function, we use local quadratic approximation coupled with a cyclic coordinate descent algorithm \cite{friedman2010regularization}. To obtain an optimizer of the GaLASSO objective function, we use a MM algorithm combined with block coordinate descent updates to handle the group LASSO component of the penalty function. The other objective functions listed in section \ref{section:methods} are special cases of SaENET(w) and GaLASSO, and therefore can be optimized with minor simplifications (see Figure \ref{flowchart}). Namely, when $\alpha=1$, SaENET(w) reduces to SaLASSO(w), and SENET(w) reduces to SLASSO(w). Furthermore, when $\hat{a}_j = 1$ for $j=1,2,..,p$, SaENET(w) reduces to SENET(w) and GaLASSO reduces to GLASSO. Optimizing the homogeneous pooled objective function with equal weights can be achieved by setting $o_i = 1/D$. 

\subsection{Optimization of SaENET(w)}

Without loss of generality, suppose that the variables in the design matrix are standardized after stacking all the imputed datasets. That is, $\sum_{d=1}^{D} \sum_{i=1}^{n} x_{d,ij} =0$, $\frac{1}{n} \sum_{d=1}^{D} \sum_{i=1}^{n} x_{d,ij}^2 = 1$, for $j=1,2,...,p$. Let $\boldsymbol{\eta} = (\boldsymbol{\eta}_1^T, \boldsymbol{\eta}_2^T,...,\boldsymbol{\eta}_D^T)^T$ be the linear predictor based on $\boldsymbol{\theta} = (\mu, \boldsymbol{\beta}^T)^T$ such that $\eta_{d,i} = \mu + x_{d,i}^T \boldsymbol{\beta}$. Then the sum of the weighted loss functions can be expressed as:
$$L(\boldsymbol{\eta}) = \frac{1}{n} \sum_{d=1}^{D} \sum_{i=1}^{n} o_i \bigg\{ -y_{d,i}\eta_{d,i}+\log\big(1+\exp(\eta_{d,i})\big) \bigg\}.$$

Let $\boldsymbol{\eta}^{(t)}$ denote the linear predictor at the $t$-th iteration. Following \citeauthor*{friedman2010regularization} \cite{friedman2010regularization}, we use a Taylor expansion at $\boldsymbol{\eta}^{(t)}$ to construct a quadratic approximation to the loss function:
$$L_Q(\boldsymbol{\eta}|\boldsymbol{\eta}^{(t)}) = \frac{1}{2n}  \sum_{d=1}^{D} \sum_{i=1}^{n} o_{i} w_{d,i} \big(\Tilde{y}_{d,i} - \mu - x_{d,i}^T \boldsymbol{\beta} \big)^2 + C\Big(\boldsymbol{\eta}^{(t)}\Big)^2,$$

\noindent where 
\begin{equation}
    \Tilde{y}_{d,i} = \mu^{(t)} + x_{d,i}^T \boldsymbol{\beta}^{(t)} + \frac{y_{d,i}-\Tilde{p}(x_{d,i})}{\Tilde{p}(x_{d,i})\big(1-\Tilde{p}(x_{d,i})\big)}
\label{SaENET_workingResponse}
\end{equation}

\noindent is the working response, $w_{d,i} = \Tilde{p}(x_{d,i})\big(1-\Tilde{p}(x_{d,i})\big)$ is a subject weight that is specific to each imputed dataset, and $$\Tilde{p}(x_{d,i}) = P(Y_{d,i} = 1 | X_{d,i} = x_{d,i}) = \bigg(1+\exp\Big(-\big(\mu^{(t)}+x_{d,i}^T \boldsymbol{\beta}^{(t)}\big)\Big)\bigg)^{-1}.$$

\noindent Going forward we will use $O_Q = L_Q\Big(\boldsymbol{\eta}|\boldsymbol{\eta}^{(t)}\Big) + \lambda P_{\alpha}(\boldsymbol{\beta})$, as shorthand notation for objective function after quadratic approximation. We then use coordinate descent to solve:
\begin{equation*}
\operatorname*{argmin}_{\boldsymbol{\theta}} O_Q = 
\operatorname*{argmin}_{\boldsymbol{\theta}} \bigg \{ 
L_Q\Big(\boldsymbol{\eta}|\boldsymbol{\eta}^{(t)}\Big) + \lambda P_{\alpha}(\boldsymbol{\beta})\bigg\}
\label{WAEnetProblem}
\end{equation*}

\noindent The derivative of the approximate objective function with respect to the intercept parameter is:
$$\frac{\partial O_Q}{\partial \mu} 
= -\frac{1}{n} \sum_{d=1}^{D} \sum_{i=1}^{n} o_i w_{d,i} \big(\Tilde{y}_{d,i} - \mu - x_{d,i}^T \boldsymbol{\beta}\big)
= -\frac{1}{n} \sum_{d=1}^{D} \sum_{i=1}^{n} o_i w_{d,i} \big(\Tilde{y}_{d,i} - x_{d,i}^T \boldsymbol{\beta}\big) + \frac{1}{n} \sum_{d=1}^{D} \sum_{i=1}^{n} o_i w_{d,i} \mu.$$

\noindent Let $z_0 = \sum_{d=1}^{D} \sum_{i=1}^{n} o_i w_{d,i} \big(\Tilde{y}_{d,i} - x_{d,i}^T \boldsymbol{\beta}\big)$. Then $\mu$ can be updated as:
\begin{equation}
    \mu^{(t+1)} \leftarrow \frac{z_0}{\sum_{d=1}^{D} \sum_{i=1}^{n} o_i w_{d,i}}
\label{SaENET_intercept}
\end{equation}
If $\beta_j > 0$,
$$\frac{\partial O_Q}{\partial \beta_j} = -\frac{1}{n} \sum_{d=1}^{D} \sum_{i=1}^{n} o_i w_{d,i}x_{d,ij} \Big(\Tilde{y}_{d,i} - \mu - x_{d, i(-j)}^T\boldsymbol{\beta}_{(-j)}\Big) + \frac{1}{n} \sum_{d=1}^{D} \sum_{i=1}^{n} o_i w_{d,i} x_{d,ij}^2 \beta_j + \lambda\alpha{\hat{a}_j} + 2 \lambda(1-\alpha) \beta_j$$
where $x_{d,i(-j)}$ refers to the value of the covariate vector for the $i$-th observation in the $d$-th imputed dataset after removing the $j$-th covariate, and $\boldsymbol{\beta}_{(-j)}$ refers to the regression coefficient vector $\boldsymbol{\beta}$ without the $j$-th entry. If $\beta_j < 0$ then the derivative of $O_Q$ with respect to $\beta_j$ is the same as the derivative when $\beta_j > 0$, the only exception being that the $\lambda\alpha{\hat{a}}_j$ term becomes $-\lambda\alpha{\hat{a}}_j$. Setting $z_{j} = \sum_{d=1}^{D} \sum_{i=1}^{n} o_i w_{d,i} x_{d,ij} \big(\Tilde{y}_{d,i} - \mu - x_{d, i(-j)}^T\boldsymbol{\beta}_{(-j)}\big) $, then $\beta_j$ can be updated as:
\begin{equation}
    \beta_j^{(t+1)} \leftarrow \frac{S \big( \frac{1}{n} z_j, \lambda\alpha{\hat{a}_j} \big) }{ \frac{1}{n} \sum_{d=1}^{D} \sum_{i=1}^{n} o_i w_{d,i} x_{d,ij}^2 + 2\lambda(1-\alpha) }
\label{SaENET_beta}
\end{equation}
where $S(z,\lambda)$ is the soft-thresholding operator:
$$S(z,\lambda) =
\left\{
	\begin{array}{ll}
		0  & \mbox{if } |z| \leq \lambda \\
		z-\lambda & \mbox{if} z>\lambda \\
		z+\lambda & \mbox{if} z<-\lambda \\
	\end{array}
\right.$$

A summary of the optimization routine for SaENET(w) is presented in Figure \ref{alg:WAEnet}. One thing to note is that we never directly compute $z_j$ after each update of $\beta_j$. A more computationally efficient approach is to update $z_j$ through the residual $r_{d,i} = \Tilde{y}_{d,i} - \mu - x_{d,i}^T\boldsymbol{\beta}$. Specifically, 
\begin{equation}
    z_{j} \leftarrow \sum_{d=1}^{D} \sum_{i=1}^{n} o_i w_{d,i} x_{d,ij} r_{d,i} + \sum_{d=1}^{D} \sum_{i=1}^{n} o_i w_{d,i} x_{d,ij}^2 \beta_j^{(t)}
\label{SaENET_z}
\end{equation}

\begin{equation*}
    r_{d,i} \leftarrow r_{d,i}' - x_{d,ij}\Big(\beta_j^{(t)} - \beta_j^{(t+1)}\Big)
\label{SaENET_r}
\end{equation*}
where $r_{d,i}'$ indicates the previous estimate of $r_{d,i}$ (we do not use the $(t)$ superscript notation here because $r_{d,i}$ is updated multiple times within the same iteration). Updating the intercept  parameter $\mu$ is similar, in that we assign $r_{d,i} \leftarrow r_{d,i}' - \big(\mu^{(t)} - \mu^{(t+1)}\big)$ and update $z_0$ accordingly. Once a coefficient is shrunk to zero, it will stay at zero for the remaining iterations.

\begin{figure}[]
    \centering
    \includegraphics[width=1\textwidth]{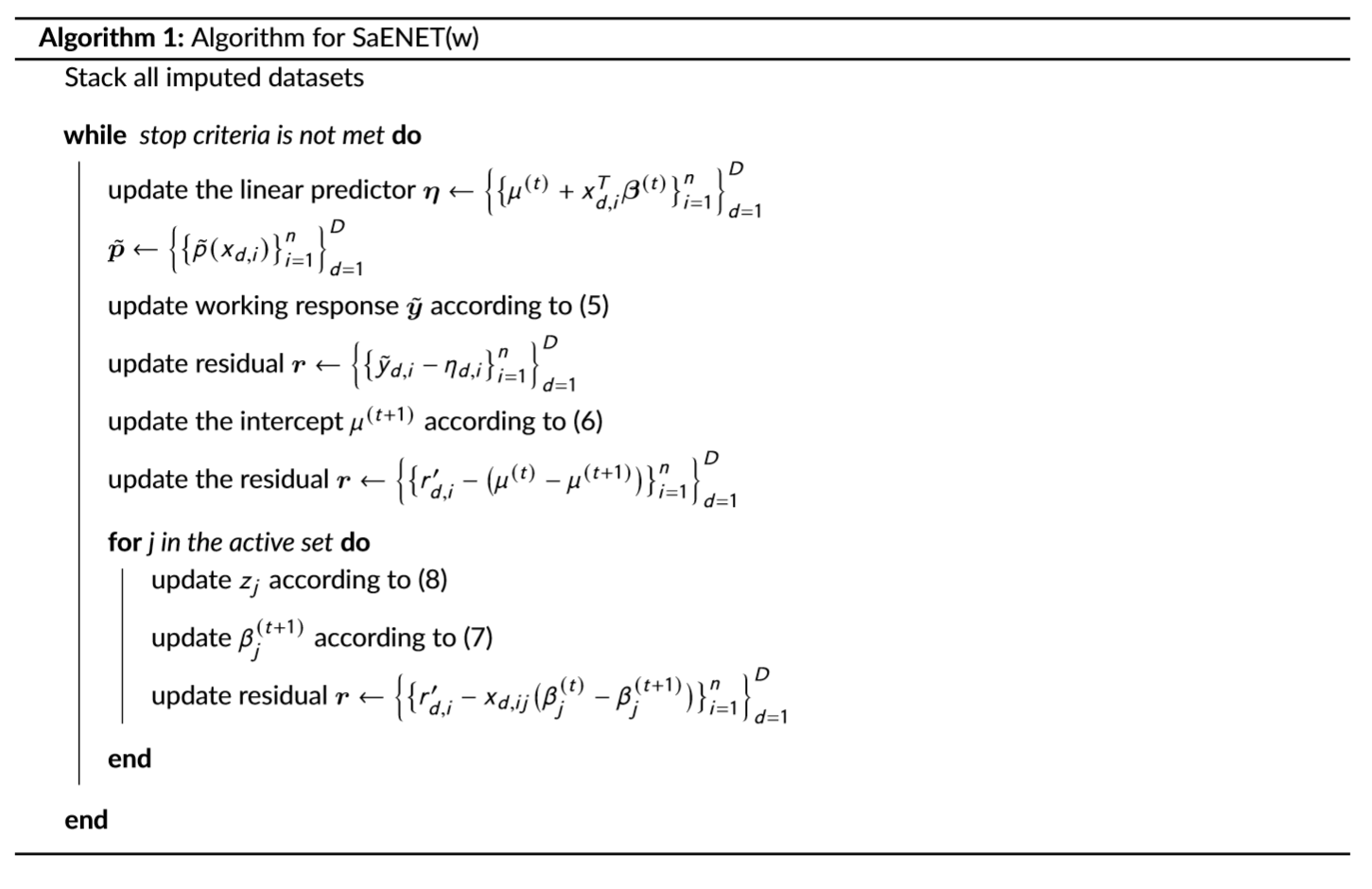}
    \caption{Algorithm illustration to optimize SaENET(w).}
    \label{alg:SaENET}
\end{figure}

\subsection{Optimization of GaLASSO}

Although \citeauthor*{chen2013variable} \cite{chen2013variable} initially proposed the idea of grouped objective functions, their optimization procedure relies on a local quadratic approximation argument to rewrite the heterogeneous pooled objective function as the sum of $D$ separate ridge regressions. Because the ridge penalty does not shrink regression coefficient estimates all the way to zero, the user is then forced to threshold the resulting regression coefficient estimates in an ad hoc manner. Another limitation of the algorithm proposed in \cite{chen2013variable} is that they only consider continuous outcome variables. In many biomedical applications, including our motivating example, the outcome is a binary indicator of disease status. To address both of these concerns, we developed a MM algorithm for GaLASSO with binary outcome data. The primary computational advantage of the MM-Algorithm is that it allows us to transform the optimization of a non-linear loss function into an optimization of a linear function, referred to as the majorizing function. Block coordinate descent can then be applied to optimize the linear majorizing function, where the blocks correspond to regression coefficient estimates for each covariate across all imputed datasets, i.e., $(\beta_{1,j}, \beta_{2,j},...,\beta_{D,j})$ for the $j$-th covariate.

Without loss of generality, we standardize the data by letting $\sum_{i=1}^{n} x_{d,ij} = 0$, $n^{-1}\sum_{i=1}^{n} x_{d,ij}^2 =1$, for $j=1,2,...,p$ and $d = 1,2,...,D$. Let $\boldsymbol{\eta} = (\boldsymbol{\eta}_1^T, \boldsymbol{\eta}_2^T,...,\boldsymbol{\eta}_D^T)^T$ be the linear predictor based on $(\boldsymbol{\theta}_1, \boldsymbol{\theta}_2,...,\boldsymbol{\theta}_D)$, where $\boldsymbol{\theta}_d = (\mu_d, \boldsymbol{\beta}_d^T)^T$, $\boldsymbol{\eta}_d = (\eta_{d,1},\eta_{d,2},...,\eta_{d,n})^T$ and $\eta_{d,i} = \mu_d + \beta_{d,1}x_{d,i1} + ... + \beta_{d,p} x_{d,ip}$. For binary outcome data, the sum of loss functions in (\ref{eq:heterogeneous_modif}) can be written as:
$$L(\boldsymbol{\eta}) = \sum_{d=1}^{D}\sum_{i=1}^{n} \Big \{ -y_{d,i} \eta_{d,i} + \log\big(1+\exp(\eta_{d,i})\big) \Big \}$$

Let $\boldsymbol{\eta}^{(t)}$ be the linear predictor at the $t$-th iteration, $L\big(\boldsymbol{\eta}^{(t)}\big)$ be the loss function given $\boldsymbol{\eta}^{(t)}$, and $\Tilde{L}\big(\boldsymbol{\eta}|\boldsymbol{\eta}^{(t)}\big)$ be the majorizing function of $L\big(\boldsymbol{\eta}^{(t)}\big)$ \cite{breheny2015group}, which can be expressed as:
$$ \Tilde{L}\big(\boldsymbol{\eta} | \boldsymbol{\eta}^{(t)}\big) = L\big(\boldsymbol{\eta}^{(t)}\big) + \big(\boldsymbol{\eta} - \boldsymbol{\eta} ^{(t)}\big) ^T \nabla L\big(\boldsymbol{\eta} ^{(t)}\big) + \frac{v}{2} \big(\boldsymbol{\eta} - \boldsymbol{\eta} ^{(t)}\big) ^T \big(\boldsymbol{\eta} - \boldsymbol{\eta} ^{(t)}\big).$$
Here, $v = max_{d} max_{i} sup_\eta \{\nabla^2 L_{d,i}(\eta) \} = 0.25$ \cite{breheny2015group}. Ignoring constants not involving $\boldsymbol{\eta}$, $\Tilde{L}\big(\boldsymbol{\eta}|\boldsymbol{\eta}^{(t)}\big)$ can be written in terms of $(\boldsymbol{\theta}_1, \boldsymbol{\theta}_2,...,\boldsymbol{\theta}_D)$:
\begin{align}
    \Tilde{L}\big(\boldsymbol{\eta}|\boldsymbol{\eta}^{(t)}\big) = \Tilde{L}(\boldsymbol{\theta}_1, \boldsymbol{\theta}_2,...,\boldsymbol{\theta}_D) \propto \frac{v}{2n} \sum_{d=1}^{D} \sum_{i=1}^{n} \big(\Tilde{y}_{d,i} - \mu_d - x_{d,i}^T\boldsymbol{\beta}_{d}\big)^2 \nonumber
\label{majorizingFunc}
\end{align}

\noindent where 
\begin{equation}
    \Tilde{y}_{d,i}=  \mu^{(t)}_{d} + x_{d,i}^T \boldsymbol{\beta}^{(t)}_d + \frac{y_{d,i}-\Tilde{p}(x_{d,i})}{v}
\label{GaLASSO_workingResponse}
\end{equation}
\noindent is the working response and $$\Tilde{p}(x_{d,i}) = P(Y_{d,i} = 1 | X_{d,i} = x_{d,i}) = \bigg(1+\exp \Big( -\Big(\mu_d^{(t)}+x_{d,i}^T {\boldsymbol{\beta}}^{(t)}_d\Big) \Big)\bigg)^{-1}.$$

Going forward, let $M_Q = \Tilde{L}(\boldsymbol{\theta}_1, \boldsymbol{\theta}_2,...,\boldsymbol{\theta}_D) + \lambda P(\boldsymbol{\beta}_1, \boldsymbol{\beta}_2,...,\boldsymbol{\beta}_D)$ denote the penalized majorizing function, which we  want to optimize:
\begin{equation*}
    \operatorname*{argmin}_{\boldsymbol{\theta}_1, \boldsymbol{\theta}_2,...,\boldsymbol{\theta}_D} M_Q = \operatorname*{argmin}_{\boldsymbol{\theta}_1, \boldsymbol{\theta}_2,...,\boldsymbol{\theta}_D} \bigg\{\Tilde{L}(\boldsymbol{\theta}_1, \boldsymbol{\theta}_2,...,\boldsymbol{\theta}_D) + \lambda P(\boldsymbol{\beta}_1, \boldsymbol{\beta}_2,...,\boldsymbol{\beta}_D) \bigg\}
\label{GAEnetEq}
\end{equation*}

To derive the block coordinate descent updates for the majorizing function optimization, we first need to introduce some new notation to distinguish coefficients within an imputed dataset from those across the imputed datasets. Let $\boldsymbol{\mu} = (\mu_1, \mu_2,...,\mu_D)^T$ be the vector of all intercept parameters across the imputed datasets, let $\boldsymbol{\beta}_{\cdot, j} = (\beta_{1,j}, \beta_{2,j},...,\beta_{D,j})^T$, for $j = 1,2,...,p$ denote the vector of the $j$-th regression coefficients across all imputed datasets, and let $\boldsymbol{\beta}_{d, \cdot} = (\beta_{d,1}, \beta_{d,2},...,\beta_{d,p})^T$ for $d = 1,2, ..., D$ indicate the regression coefficients for the $d$-th imputed dataset. If we take the subdifferential of $M_Q$ with respect to $\boldsymbol{\mu}$, we get: $$\frac{\partial M_Q}{\partial \boldsymbol{\mu}} = -\frac{v}{n} \boldsymbol{z}_0 + v \boldsymbol{\mu}$$
\noindent where 
$$
\boldsymbol{z}_0 = 
\left[\begin{array}{c}
\sum_{i=1}^{n} \big(\Tilde{y}_{1,i} - x_{1,i}^T \boldsymbol{\beta}_{1, \cdot}\big)\\
\sum_{i=1}^{n} \big(\Tilde{y}_{2,i} - x_{2,i}^T \boldsymbol{\beta}_{2, \cdot}\big)\\
\vdots\\
\sum_{i=1}^{n} \big(\Tilde{y}_{D,i} - x_{D,i}^T \boldsymbol{\beta}_{D, \cdot}\big)\\
\end{array}\right]
$$ \noindent Since $x_{d,ij}$ have been centered, the intercept can be updated as:
\begin{equation}
    \boldsymbol{\mu}^{(t+1)} \leftarrow 
    \left[\begin{array}{c}
    \frac{1}{n} \sum_{i=1}^{n} \Tilde{y}_{1,i}\\
    \frac{1}{n} \sum_{i=1}^{n} \Tilde{y}_{2,i}\\
    \vdots\\
    \frac{1}{n} \sum_{i=1}^{n} \Tilde{y}_{D,i}\\
    \end{array}\right]
\label{GaLASSO_intercept}
\end{equation}

We now derive the update for the block of coefficients $\boldsymbol{\beta}_{\cdot, j}$, $j=1,...,p$. If $\boldsymbol{\beta}_{\cdot,j} \neq \boldsymbol{0}$, the subdifferential of $M_Q$ with respect to $\boldsymbol{\beta}_{\cdot,j}$ is: $$\frac{\partial M_Q}{\partial \boldsymbol{\beta}_{\cdot, j}} = -\frac{v}{n} \boldsymbol{z}_j + v\boldsymbol{\beta}_{\cdot, j} + \lambda{\hat{a}_j} \frac{\boldsymbol{\beta}_{\cdot,j}}{||\boldsymbol{\beta}_{\cdot,j}||}
\label{subdifferential2}
$$ where $$
\boldsymbol{z}_j =
\left[\begin{array}{c}
\sum_{i=1}^{n} x_{1,ij}\big(\Tilde{y}_{1,i} - \mu_1 - x_{1,i(-j)}^T \boldsymbol{\beta}_{1,(-j)}\big) \\
\sum_{i=1}^{n} x_{2,ij}\big(\Tilde{y}_{2,i} - \mu_2 - x_{2,i(-j)}^T \boldsymbol{\beta}_{2,(-j)}\big) \\
\vdots\\
\sum_{i=1}^{n} x_{D,ij}\big(\Tilde{y}_{D,i} - \mu_D - x_{D,i(-j)}^T \boldsymbol{\beta}_{D,(-j)}\big) \\
\end{array}\right]
$$

Here, $x_{d,i(-j)}$ is the vector of covariates for the $i$-th observation in the $d$-th dataset after removing $x_{d,ij}$, and $\boldsymbol{\beta}_{d,(-j)}$ is the corresponding vector of regression coefficients. The form of the partial derivative with respect to $\boldsymbol{\beta}_{\cdot, j}$, indicates that the $\boldsymbol{\beta}_{\cdot, j}$ update must lie on the line segment joining the zero vector, $\boldsymbol{0}$, and $\boldsymbol{z}_j$. Therefore, $\boldsymbol{\beta}_{\cdot, j}$ can be updated as:
\begin{equation}
    \boldsymbol{\beta}_{\cdot, j}^{(t+1)} \leftarrow 
    \frac{1}{v} S \bigg(\frac{v}{n} ||\boldsymbol{z}_j|| , \lambda{\hat{a}_j} \bigg) \frac{\boldsymbol{z}_j}{||\boldsymbol{z}_j||}
\label{GaLASSO_beta}
\end{equation}

\begin{figure}[]
    \centering
    \includegraphics[width=1\textwidth]{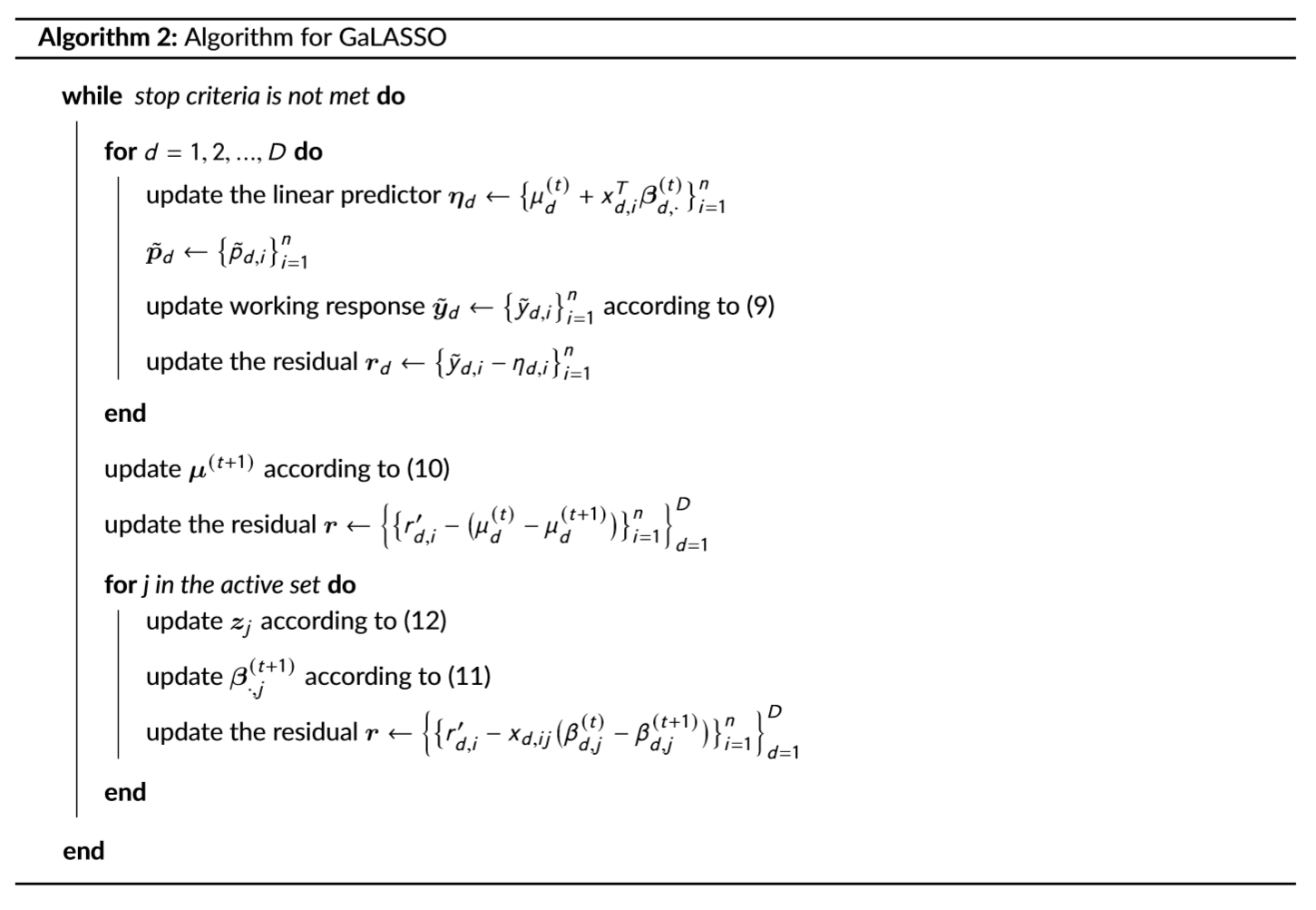}
    \caption{Algorithm illustration to optimize GaLASSO.}
    \label{alg:GaLASSO}
\end{figure}

As with SaENET(w), we do not directly calculate $\boldsymbol{z}_j$ but use the residual $r_{d,i}$ to update $\boldsymbol{z}_j$. Let $r_{d,i} = \Tilde{y}_{d,i} - \mu_d - x_{d,i}^T \boldsymbol{\beta}_{d,\cdot}$. Then 
\begin{equation}
    \boldsymbol{z}_j \leftarrow
    \left[\begin{array}{c}
    z_{1,j}\\
    z_{2,j}\\
    \vdots\\
    z_{D,j}\\
    \end{array}\right] =
    \left[\begin{array}{c}
    \sum_{i=1}^{n} x_{1,ij}r_{1,i} + n \beta_{1,j}^{(t)}\\
    \sum_{i=1}^{n} x_{2,ij}r_{2,i} + n \beta_{2,j}^{(t)}\\
    \vdots\\
    \sum_{i=1}^{n} x_{D,ij}r_{D,i} + n \beta_{D,j}^{(t)}\\
\end{array}\right]
\label{GaLASSO_z}
\end{equation}
$$r_{d,i} \leftarrow r_{d,i}' - x_{d,ij}\Big(\beta_{d,j}^{(t)} - \beta_{d,j}^{(t+1)}\Big) $$ 
Updating the intercept  parameter $\boldsymbol{\mu}$ is similar, in that we assign $r_{d,i} \leftarrow r_{d,i}' - \big(\mu_d^{(t)} - \mu_d^{(t+1)}\big)$ and update $z_0$ accordingly. Once the block of regression coefficients corresponding to the $j$-th covariate is shrunk to zero, it will stay at zero for the remaining iterations. A summary of the optimization routine is described in Figure \ref{alg:GaLASSO}.

\subsection{Tuning parameters}

The implementation of SaENET(w) and GaLASSO in the \textit{miselect} package sets tuning parameter values based on a 5-fold cross-validation routine combined with a "one-standard-error" rule \cite{hastie2009elements, hastie2015statistical}. More specifically, GaLASSO chooses the largest $\lambda$, such that the cross-validation error is within one standard error of the minimum cross-validation error. SaENET(w) selects an $(\alpha,\lambda)$ pair by first identifying all dyads whose cross-validation errors are within one standard error of the minimum cross-validation error, and then selecting the $(\alpha,\lambda)$ pair with the largest $L_1$ penalty, i.e, $\lambda\alpha$. For both GaLASSO and SaENET(w), the default sequence for $\lambda$ ranges from $\lambda_{min}$ to $\lambda_{max}$ on the log scale where $\lambda_{max}$ is chosen to be the smallest value where all the coefficients are shrunk to zero, and $\lambda_{min} = 10^{-6}\lambda_{max}$. Since $\alpha \in [0,1]$, then for SaENET(w) we can choose a sequence of values, say $(0,0.1,0.2,...,0.9,1)$, to fully explore the tradeoff between $L_1$ and $L_2$ shrinkage. When the non-adaptive methods are used, i.e. GLASSO, SENET(w), SLASSO(w), we set $\lambda_{min} = 10^{-3} \lambda_{max}$. The reason that $\lambda_{min}$ is smaller for the adaptive methods is to prevent large adaptive weights from overwhelming the overall shrinkage.

One caveat with cross-validation to select tuning parameters for the stacked objective function approach is that, by stacking the imputed datasets on top of one another, there are now $D$ rows corresponding to the same subject. Therefore, if the $D$ rows corresponding to the same subject are distributed across the validation folds, then we are prone to overfitting. To prevent this issue, we restrict the fold assignment such that all of the rows corresponding to the same subject are assigned to a single validation fold.

\section{Data Example}

Our motivating example is derived from data in the University of Michigan ALS Patient Repository which aims to identify environmental risk factors associated with ALS \cite{su2016association, goutman2019high, yu2014environmental}. ALS is progressive disease primarily involving motor neuron cells in the brain and spinal cord leading to weakness of voluntary muscles and death within 2-4 years due to respiratory failure \cite{goutman2017diagnosis}. ALS has a complex etiology driven by the combination of genetic susceptibility and environmental exposures \cite{paez2015amyotrophic, goutman2018emerging, al2013epidemiology}. In this particular study we are interested in characterizing the relationship between persistent organic pollutant (POP) exposure and ALS susceptibility. In total, 167 ALS cases and 99 healthy controls were recruited between 2011 and 2014. All participants provided written informed consent and the study was IRB approved \cite{su2016association, goutman2019high}. Participants provided their weight and height measurements at the time of study enrollment and 5 years prior to enrollment and their educational attainment. Plasma samples were collected from each study participant to measure 122 potentially neurotoxic POPs, which can be broadly partitioned into three chemical classes: organochlorine pesticides, polychlorinated biphenyls (PCBs), and polybrominated diphenyl ethers (PBDEs). Of the 122 POPs, a subset of 23 POPs with less than 60\% non-detects were used for further analysis. More information regarding data collection and study protocols can be found in \cite{su2016association, goutman2019high}.

From a statistical perspective, the target model of interest is a penalized logistic regression model, where the outcome is ALS case/control status and the design matrix of predictors and covariates contains the 23 log-transformed POPs, age, sex, body mass index (BMI), rate of BMI change over the five years prior to survey consent, and education. The confounders are selected based on existing ALS literature \cite{chio2009prognostic}. Elastic net regularization is of particular interest here, because many of the POPs have medium pairwise correlations with one another (Figure S1 in the supplementary materials) \cite{goutman2019high}. We only penalize regression coefficients corresponding to POPs to ensure that adjustment covariates are retained in the final model. If we look at the percent missingness (Table S1 in the supplementary materials), we observe that 9 of the 23 variables have more than 30\% below the detection limit and 24.4\% of subjects have incomplete BMI. Complete-case analysis in this context is infeasible, as every control is missing at least one covariate or has at least one measured POP below its respective detection limit. POP concentrations below their respective detection limits were imputed 50 times conditional on case/control status following the censored likelihood multiple imputation strategy outlined in \cite{boss2019estimating}. After imputing the exposure non-detects, multiple imputation by chained equations (MICE) was used to impute the missing adjustment covariates \cite{vanbuuren2011}.

\begin{table}[h]
\scriptsize
\centering
\caption{\label{adhocResults} The proportion of imputed datasets in which each POP is selected. Note that only 11 out of the 23 POPs are listed because the other 12 POPs were never selected by LASSO, aLASSO, ENET, and aENET. Bolded entries indicate a selection proportion over 0.50 and dashes denote a selection proportion of zero. D is the number of imputed datasets. The total number of POPs in the final active set, as determined by three ad hoc combining rules, are presented in the rows titled, Union (in at least one dataset), 50\%-cutoff (in over 50\% datasets), and Intersection (in all datasets).}
\begin{tabular}{p{1.5cm}p{1.1cm}p{1.1cm}p{1.1cm}p{1.1cm}p{1.1cm}p{1.1cm}p{1.1cm}p{1.1cm}}
\hline
\multirow{2}{*}{POPs} & \multicolumn{2}{c}{LASSO} & \multicolumn{2}{c}{aLASSO} & \multicolumn{2}{c}{ENET} & \multicolumn{2}{c}{aENET} \\
 & D=10 & D=50 & D=10 & D=50 & D=10 & D=50 & D=10 & D=50 \\ \hline
PBDE 28 & - & 0.04 & - & - & 0.10 & 0.14 & - & - \\
PBDE 99 & - & - & - & - & 0.10 & 0.08 & - & - \\
PBDE 153 & \textbf{1.00} & \textbf{1.00} & - & 0.02 & \textbf{1.00} & \textbf{1.00} & - & 0.02 \\
PeCB & \textbf{1.00} & \textbf{1.00} & \textbf{1.00} & \textbf{1.00} & \textbf{1.00} & \textbf{1.00} & \textbf{1.00} & \textbf{1.00} \\
trans-chlordane & \textbf{1.00} & \textbf{0.94} & - & 0.04 & \textbf{1.00} & \textbf{1.00} & - & 0.04 \\
cis-nonachlor & \textbf{1.00} & \textbf{1.00} & \textbf{1.00} & \textbf{0.96} & \textbf{1.00} & \textbf{1.00} & \textbf{1.00} & \textbf{0.96} \\
PCB 110 & - & - & - & - & 0.10 & 0.06 & - & - \\
PCB 151 & \textbf{1.00} & \textbf{0.98} & - & - & \textbf{1.00} & \textbf{1.00} & - & - \\
PCB 174 & 0.10 & 0.02 & - & - & 0.10 & 0.14 & - & - \\
PCB 180 & - & - & - & - & 0.10 & 0.02 & - & - \\
PCB 202 & 0.10 & 0.02 & - & - & 0.10 & 0.08 & - & - \\
\hline
\textbf{Union} & 7 & 8 & 2 & 4 & 11 &  11 & 2 & 4 \\
\textbf{50\%-cutoff} & 5 & 5 & 2 & 2 & 5 & 5 & 2 & 2 \\
\textbf{Intersection} & 5 & 3 & 2 & 1 & 5 & 5 & 2 & 1 \\
\hline
\end{tabular}
\end{table}


To illustrate the problem with fitting separate penalized regression routines, we first apply LASSO, aLASSO, ENET, and aENET to each imputed dataset and calculate the proportion of imputed datasets in which each variable is selected. In Table \ref{adhocResults}, we summarize the results for scenarios with 10 and 50 imputed datasets. The final active set for each method can be determined using the ad hoc rules outlined in \citeauthor{wood2008} \cite{wood2008}, namely (i) a variable is considered selected if it is selected in all imputed datasets, (ii) a variable is considered selected if it is selected in at least one imputed dataset, (iii) a variable is considered selected if it is selected in at least half of the imputed datasets. Note that depending on whether we use (i), (ii), or (iii), the number of selected POPs changes, especially when $D = 50$. For example, if ad hoc combining rule (i) is used, then ENET selects PBDE 153, pentachlorobenzene (PeCB), trans-chlordane, cis-nonachlor, and PCB 151. However, ad hoc combining rule (ii) additionally selects of PBDE 28, PBDE 99, PCB 110, PCB 174, PCB 180, and PCB 202. As expected, the final active set determined by aLASSO and aENET is much more sparse than their non-adaptive counterparts; if ad hoc combining rule (iii) is used then aLASSO and aENET only select PeCB and cis-nonachlor.

A more subtle point that deserves further comment is that non-uniform POPs selection across imputed datasets makes it difficult to obtain final regression coefficient estimates. For example, consider aLASSO when $D = 50$, which selects PeCB in all 50 imputed datasets. Although PeCB is always selected, the interpretation of the regression coefficient for PeCB is conditional on the other selected exposures. That is, despite that PeCB is selected for all imputed datasets, the PeCB regression coefficient estimates are not necessarily comparable across imputed datasets because they condition on different sets of selected exposures.

\begin{table}[h]
\scriptsize
\centering
\caption{\label{table:realDataAppD50} Regression coefficient estimates for six POPs collected as part of the University of Michigan ALS Clinic case-control study (167 ALS cases and 99 healthy controls). Results are based on 50 imputed datasets. Only six of the 23 POPs are displayed because the other 17 were not selected by any method.}
\begin{tabular}{p{1.2cm}p{0.8cm}p{0.9cm}p{0.8cm}p{0.8cm}p{0.8cm}p{0.9cm}p{0.8cm}p{0.8cm}p{0.8cm}p{1cm}}
\hline
POPs & SLASSO & SaLASSO & SENET & SaENET & \begin{tabular}[c]{@{}c@{}}SLASSO\\ (w)\end{tabular} & \begin{tabular}[c]{@{}c@{}}SaLASSO\\ (w)\end{tabular} & \begin{tabular}[c]{@{}c@{}}SENET\\ (w)\end{tabular} & \begin{tabular}[c]{@{}c@{}}SaENET\\ (w)\end{tabular} & GLASSO & GaLASSO \\ \hline
PBDE 28 & - & - & - & - & - & - & 0.007 & - & - & - \\
PBDE 153 & 0.085 & - & 0.093 & - & 0.086 & 0.013 & 0.109 & 0.015 & - & - \\
PeCB & 0.329 & 0.716 & 0.383 & 0.697 & 0.261 & 0.675 & 0.338 & 0.558 & 0.415 & 0.788 \\
trans-chlordane & 0.050 & - & 0.111 & - & 0.058 & - & 0.116 & - & 0.076 & - \\
cis-nonachlor & 0.196 & 0.527 & 0.310 & 0.529 & 0.226 & 0.533 & 0.295 & 0.458 & 0.303 & 0.567 \\
PCB 151 & \textbf{-} & 0.177 & 0.143 & 0.196 & - & - & - & - & 0.183 & 0.097 \\
\textbf{\#selected} & \textbf{4} & \textbf{3} & \textbf{5} & \textbf{3} & \textbf{4} & \textbf{3} & \textbf{5} & \textbf{3} & \textbf{4} & \textbf{3} \\
\textbf{\#removed} & \textbf{19} & \textbf{20} & \textbf{18} & \textbf{20} & \textbf{19} & \textbf{20} & \textbf{18} & \textbf{20} & \textbf{19} & \textbf{20} \\ \hline
\end{tabular}
\end{table}

The regression coefficient estimates obtained from the stacked and grouped pooled objective function methods with 50 imputed datasets are in Table \ref{table:realDataAppD50} and with 10 imputed datasets are in Table S2 in the supplementary materials. Because the grouping methods produce different regression coefficient estimates across imputed datasets, the final estimates presented in Table \ref{table:realDataAppD50} are the average of the regression coefficient estimates across imputed datasets. All coefficient estimates are positive, which is consistent with the hypothesis that higher POP exposure positively associates with a higher ALS risk. All of the non-adaptive methods select PeCB, trans-chlordane, and cis-nonachlor, and all of the non-adaptive variants of the stacked objective functions also select PBDE 153. Similarly, the adaptive methods all select PeCB and cis-nonachlor, however the unweighted stacked objective functions and the grouped objective functions additionally select PCB 151 while the weighted stacked objective functions additionally select PBDE 153. Since all methods select PeCB and cis-nonachlor, we conclude that further studies should be conducted to assess the PeCB and cis-nonachlor neurotoxicity.

\section{Simulation}

We now evaluate the performance of the stacked methods and the grouped methods mentioned in Section 2, including SLASSO, SaLASSO, SENET, SaENET, SLASSO(w), SaLASSO(w), SENET(w), SaENET(w), GLASSO, and GaLASSO.

\subsection{Simulation setting}

We simulate 1000 datasets of size $n$ containing outcome $Y$ and $p$ covariates $\boldsymbol{X}$ under 4 different cases. For Cases 1 and 2 we take $n$=500 and $p$=20, and for Cases 3 and 4 we take $n=1000$ and $p=100$. In all cases, covariates are generated from a multivariate normal distribution with zero mean and unit variance. The correlation structure of the covariates is block-diagonal, in order to mimic the correlation structure of the POPs. A more comprehensive breakdown of the correlation structure is detailed in Table \ref{table1:simulation_cases}.

Given $\boldsymbol{X}$, we generate a binary $Y$ from 
\begin{center}
$logit(P(Y = 1 \vert X)) = \beta_0 + \beta_1 X_1 + ... + \beta_{p} X_{p}$
\end{center}
The true value of $\boldsymbol{\beta}$ is specified according to different simulation cases, where Cases 1 and 3 correspond to concentrated signals and Cases 2 and 4 correspond to distributed signals. Here, signals are concentrated if there is only one non-null coefficient in a group, and signals are distributed if there is more than one non-null coefficient in a group. Regression coefficient magnitudes are set to fix the prevalence of $Y = 1$ at about 50\% and maintain the Cox-Snell pseudo-R$^2$ at approximately 0.5. The four simulation settings are summarized in Table \ref{table1:simulation_cases}.

Missing values are generated under Missing at Random (MAR) assumption \cite{little2019statistical}. In all cases the outcome and the last covariate $X_p$ are fully observed and the missingness indicator $R_{j}$ for covariate $X_j$ is generated from the logistic regression model
$$logit(Pr(R_{j}=1)) = \alpha_{0j} + \alpha_{1j}X_{p} + \alpha_{2j} Y$$

\noindent where $R_{j} = 1$ indicates that covariate $X_j$ is missing, and $\alpha_{0j}, \alpha_{1j}$ and $\alpha_{2j}$ are chosen to control the percentage of missingness for $X_j$. For Case 1 and Case 2, about 25\% of subjects are missing $\{X_{1},..., X_{5}\}$, 35\% are missing $\{X_{6},...,X_{13}\}$, 45\% are missing $\{X_{14},...,X_{17}\}$ and 55\% are missing $\{ X_{18}, X_{19}\}$. For Case 3 and Case 4, about 25\% of subjects are missing $\{X_1,...,X_{30}\}$, 35\% are missing $\{X_{31},...,X_{60}\}$, 45\% are missing  $\{X_{61},...,X_{82}\}$, 55\% are missing $\{X_{83},...,X_{95} \}$, and 60\% are missing $\{X_{96},...,X_{99} \}$. In total, less than 5\% of subjects have complete data, and about 13\% subjects have more than 90\% data for all covariates. R package \textit{mice} \cite{zhang2016multiple} is used to multiply impute the missing data using predictive mean matching. Each simulated dataset is imputed 10 times and the stacked and grouped methods are then applied to perform variable selection and parameter estimation.

\begin{table}[h]
\scriptsize
\caption{\label{tab:table-name} Data generation details for all simulation settings. In the correlation structure column, covariates within the same parentheses have an exchangeable pairwise correlation structure with one another, but are independent from all other covariates. For Case 1 and Case 2, $n=500$, $p=20$, approximately 25\% subjects are missing $\{X_{1},..., X_{5}\}$, 35\% are missing $\{X_{6},...,X_{13}\}$, 45\% are missing $\{X_{14},...,X_{17}\}$, and 55\% are missing $\{ X_{18}, X_{19}\}$. For Case 3 and Case 4, $n=1000$, $p=100$, about 25\% subjects are missing $\{X_1,...,X_{30}\}$, 35\% are missing $\{X_{31},...,X_{60}\}$, 45\% are missing $\{X_{61},...,X_{82}\}$, 55\% are missing $\{X_{83},...,X_{95} \}$, and 60\% are missing $\{X_{96},...,X_{99} \}$. The No. Signals (\%) column refers to the number of covariates that have non-null coefficients and its percentage.}
 
\begin{tabular}{p{0.9cm}p{2.8cm}p{2cm}p{2.2cm}p{4.2cm}}

\hline
\textbf{}      & \textbf{Correlation structure} & \textbf{Signal structure} & \textbf{No. Signals (\%)} & \textbf{Signals}\\
\hline
\textbf{Case 1} &  \begin{tabular}[c]{@{}l@{}} $(X_1,X_2,X_3)$ as 0.9\\     $(X_6, X_7,X_8)$ as 0.5\\     $(X_{11}, X_{12}, X_{13})$ as 0.3\end{tabular}       & Concentrated     & 5  (25\%)                               & \begin{tabular}[c]{@{}l@{}}$\beta_1 = 2,     \beta_4 = 1.5,     \beta_7 = 1.5$,\\    $\beta_{11} = 1,    \beta_{14} = 1$;\end{tabular}                                                                               \\
\textbf{Case 2} &\begin{tabular}[c]{@{}l@{}} $(X_1,X_2,X_3)$ as 0.9\\     $(X_6, X_7,X_8)$ as 0.5\\     $(X_{11}, X_{12}, X_{13})$ as 0.3\end{tabular}          & Distributed       & 5  (25\%)                               & \begin{tabular}[c]{@{}l@{}}$\beta_1 = 2, \beta_2 = 1,    \beta_4 = 2$,\\     $\beta_7 = 1,     \beta_{11} = 1$\end{tabular}                                                                                       \\
\textbf{Case 3} &\begin{tabular}[c]{@{}l@{}} $(X_1,..., X_6)$ as 0.9\\     $(X_{11}, ...,X_{16})$ as 0.5\\     $(X_{21}, ..., X_{26})$ as 0.3\end{tabular}        & Concentrated      & 10  (10\%)                               & \begin{tabular}[c]{@{}l@{}}$\beta_2 = 2,    \beta_7 = 0.8,    \beta_9 = 0.8,    \beta_{12} = 0.5$,\\     $\beta_{17} = 1.5,    \beta_{27} = 1,     \beta_{37} = 0.8$,    \\ $\beta_{47} = 0.4,     \beta_{48} = 1,     \beta_{49} = 1$;\end{tabular} \\
\textbf{Case 4} &\begin{tabular}[c]{@{}l@{}} $(X_1,..., X_6)$ as 0.9\\     $(X_{11}, ...,X_{16})$ as 0.5\\     $(X_{21},..., X_{26})$ as 0.3\end{tabular}         & Distributed       & 10  (10\%)                               & \begin{tabular}[c]{@{}l@{}}$\beta_{1} = 1.2, \beta_{2} = 0.8, \beta_{3} = 0.4, \beta_{4} = 0.4$,\\     $\beta_{12} = 1.2, \beta_{13} = 1,     \beta_{17} = 1.2$, \\ $\beta_{27} = 1,  \beta_{37}=1, \beta_{47} = 1$; \end{tabular} \\ \hline

\end{tabular}
\label{table1:simulation_cases}
\end{table}

\subsection{Simulation results}

We evaluate simulation results in terms of the following four metrics. In the following definitions, T and F are the number of non-null and null coefficients in the data generating model, respectively, R is the total number of simulation runs, $\hat{\beta}_{j}^{r}$ is the coefficient estimate for $\beta_j$ in the $r$-th run.
Since the estimates for the $j$-th coefficient by GLASSO and GaLASSO are different across imputed datasets, the mean $\frac{1}{D} \sum_{d=1}^{D} \hat{\beta}_{d,j}^{r}$ is used to approximate $\hat{\beta}_j^{r}$.

\begin{itemize}
    \item Sensitivity (SENS) $ = \frac{1}{RT} \sum_{r=1}^{R}\bigg( \#\ of\ selected\ non-null\ coefficients\ in\ the\ r_{th}\ run\bigg)$
    \item Specificity (SPEC)$ = \frac{1}{RF} \sum_{r=1}^{R}\bigg( \#\ of\ selected\ null\ coefficients\ in\ the\ r_{th}\ run \bigg)$
    \item Mean squared error for non-null coefficients (MSE\textsubscript{non-null})$ = \frac{1}{R} \sum_{r=1}^{R} \sum_{j=1}^{p} (\hat{\beta}_j^{r} - \beta_j)^2 I\bigg( \beta_j\ is\ a\ non-null\ coefficient\bigg)$
    \item Mean squared error for null coefficients (MSE\textsubscript{null}) $ = \frac{1}{R} \sum_{r=1}^{R} \sum_{j=1}^{p} (\hat{\beta}_j^{r} - \beta_j)^2 I\bigg( \beta_j\ is\ a\ null\ coefficient\bigg)$
\end{itemize}

\noindent Sensitivity and specificity capture the accuracy of variable selection and vary between 0 and 1, with larger values indicating a better performance. The MSEs capture the estimate accuracy and smaller values indicate a better performance.

\begin{figure}[h]
    \centering
    \includegraphics[width=1\textwidth]{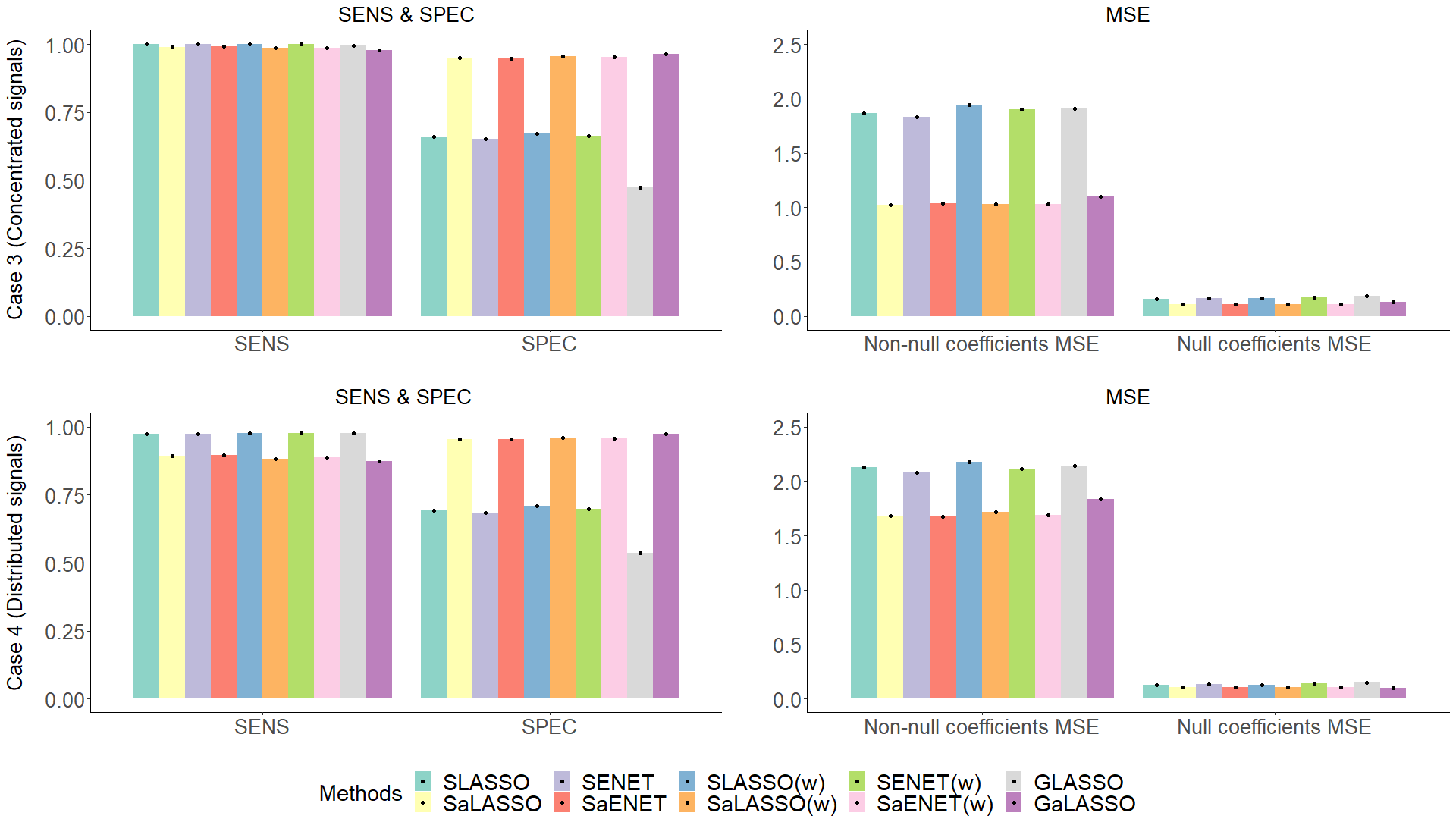}
    \caption{Simulation results for Case 1 (top panel) and Case 2 (bottom panel) where n=500 and p=20 for 10 imputed datasets. Sensitivity (SENS) and specificity (SPEC) are on the left and MSE for non-null and null coefficients are on the right.}
    \label{fig:case1&2_D10}
\end{figure}

\begin{figure}[h]
    \centering
    \includegraphics[width=1\textwidth]{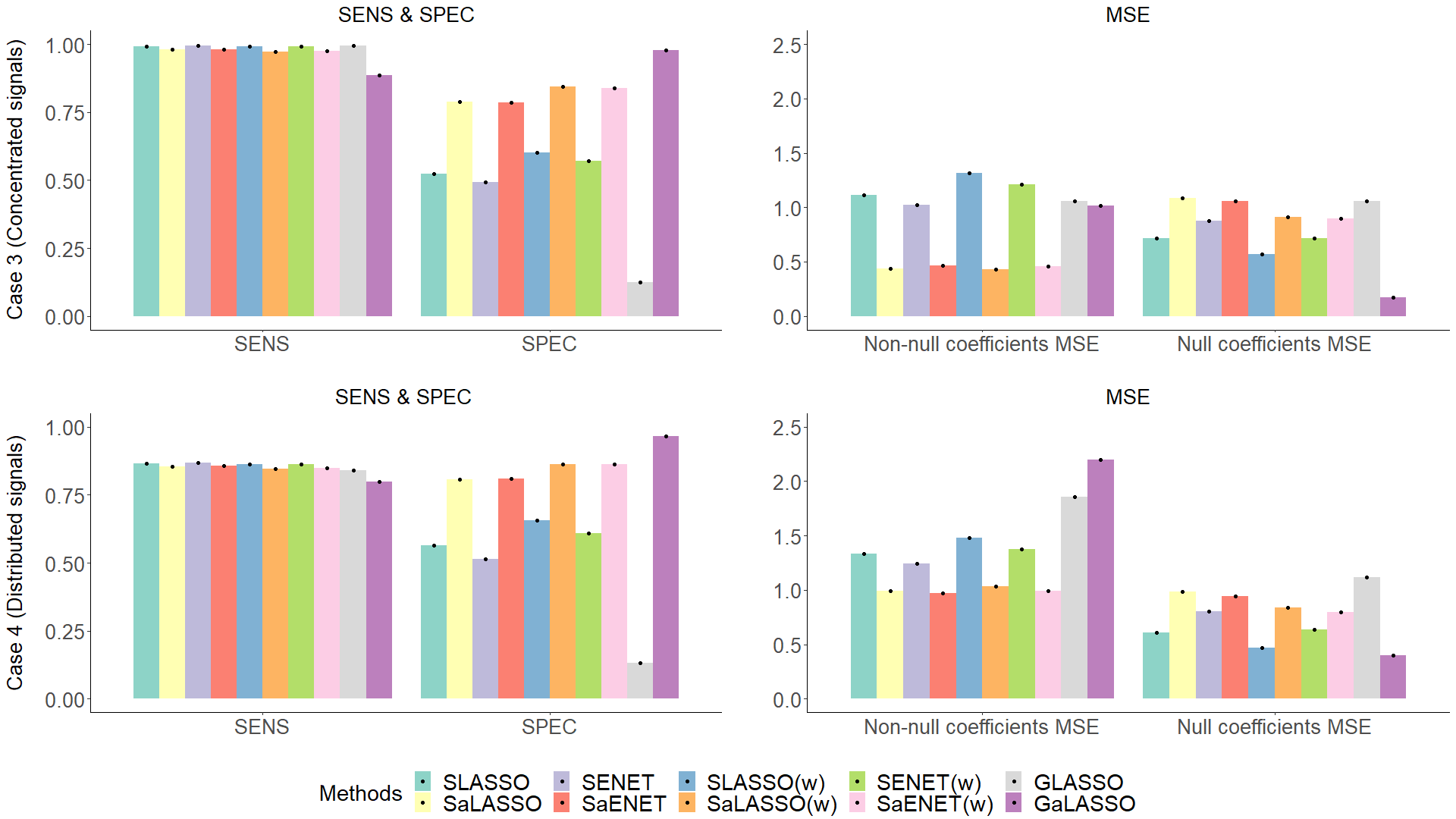}
    \caption{Simulation results for Case 3 (top panel) and Case 4 (bottom panel) where n=1000 and p=100 for 10 imputed datasets. Sensitivity (SENS) and specificity (SPEC) are on the left and MSE for non-null and null coefficients are on the right.}
    \label{fig:case3&4_D10}
\end{figure}

Figure \ref{fig:case1&2_D10} and Figure \ref{fig:case3&4_D10} present sensitivity, specificity, MSE for non-null coefficients, and MSE for null coefficients for all four cases. Overall, compared to the corresponding non-adaptive methods, the adaptive methods perform better with respect to estimation and selection. Specifically, the adaptive methods have similar sensitivity but considerably higher specificity, and considerably smaller MSE for non-null coefficients, except for GaLASSO and GLASSO under Case (Figure 3). For null coefficients, the adaptive methods have similar MSE to those of non-adaptive methods under Cases 1 and 2, and noticeably larger MSE under Cases 3 and 4, except for GaLASSO and GLASSO under Case 4. Adaptive methods have high sensitivity across all cases and high specificity under Cases 1 and 2. Under Cases 3 and 4, GaLASSO has the highest specificity, followed by SaLASSO(w) and SaENET(w). All adaptive methods have similar MSE under Cases 1 and 2 for both non-null and null coefficients. On the other hand, under Cases 3 and 4, GaLASSO has the largest MSE for non-null coefficients and the smallest MSE for null coefficients, which is due to relatively low sensitivity and high specificity compared to other adaptive methods.

The average runtime for each method is presented in Table \ref{timeSpent}. Compared to the grouped methods, the stacked methods are faster in all cases. Speed of SLASSO is 13.8 times faster than GLASSO, and SaLASSO is 2.8 times faster than GaLASSO in Case 1 when the sample size is small. When the sample size is large, i.e. Case 3, SLASSO is 15.7 times faster than GLASSO, and SaLASSO is 2.3 times faster than GaLASSO. The adaptive methods have longer average runtimes than the non-adaptive methods because the adaptive methods require an elastic net initialization step to construct the adaptive weights.

\begin{table}[h]
\scriptsize
\caption{Average runtime (minutes) for each method for 4 cases with 10 imputed datasets. Case 1 and Case 2 have 500 observations and 20 covariates. Case 3 and Case 4 have 1000 observations and 100 covariates.}
\label{timeSpent}
\begin{tabular}{p{1.2cm}p{0.8cm}p{0.9cm}p{0.8cm}p{0.8cm}p{0.8cm}p{0.9cm}p{0.8cm}p{0.8cm}p{0.8cm}p{1cm}}
\hline
 & SLASSO & SaLASSO & SENET & SaENET & \begin{tabular}[c]{@{}c@{}}SLASSO\\ (w)\end{tabular} & \begin{tabular}[c]{@{}c@{}}SaLASSO\\ (w)\end{tabular} & \begin{tabular}[c]{@{}c@{}}SENET\\ (w)\end{tabular} & \begin{tabular}[c]{@{}c@{}}SaENET\\ (w)\end{tabular} & GLASSO & GaLASSO \\
\hline
Case1 & 1.87 & 9.30 & 8.59 & 11.95 & 1.76 & 8.65 & 7.95 & 11.23 & 27.64 & 35.21 \\
Case2 & 2.07 & 9.16 & 8.28 & 12.66 & 1.95 & 8.52 & 7.65 & 11.72 & 35.25 & 45.31 \\
Case3 & 29.95 & 160.72 & 153.88 & 189.87 & 32.20 & 170.46 & 163.11 & 200.46 & 500.12 & 524.38 \\
Case4 & 26.31 & 141.72 & 132.36 & 180.46 & 21.87 & 113.89 & 105.95 & 144.24 & 422.60 & 459.17\\
\hline
\end{tabular}
\end{table}

\section{Discussion}

In this paper, we elucidated the difference between stacked and grouped pooled objective functions, which are both designed to achieve uniform variable selection across multiple imputed datasets. The stacked pooled objective function assumes that the underlying true signals are the same across imputed datasets, including the signal magnitude, while the grouped pooled objective function assumes uniform signal selection but allows for different active signal magnitudes across imputed datasets. We extended existing methods to handle binary outcomes, developed a MM algorithm combined with block coordinate descent updates to optimize grouped pooled objective functions for LASSO and aLASSO regularization, and derived cyclic coordinate descent algorithm for the stacked pooled objective functions with ENET and aENET regularization. Algorithms for implementing the stacked and grouped approaches outlined in Section \ref{section:methods}, for both continuous and binary outcomes, are available in the \textit{miselect} R package.

From a practical perspective, there are several reasons that one might prefer stacked over grouped pooled objective functions. Based on our simulations, the overall MSE for estimates generated by optimizing stacked pooled objective functions was either smaller than or equal to the estimates generated by optimizing the grouped pooled objective function, provided that adaptive weighting is used. We also observed that the total runtime for optimizing stacked pooled objective functions is noticeably lower compared to the grouped pooled objective function optimization routine. Moreover, the stacked pooled objective functions are much easier to extend beyond ENET penalization. For example, if one wanted to use a hierarchical interaction detection penalty, such as hierNet \cite{bien2013}, one would only need to download the \textit{hierNet} package from CRAN, and use the existing hierNet implementation on the stacked imputed datasets. Conversely, grouped methods would necessitate developing of additional algorithms to optimize an objective function with both a group lasso penalty and a hierarchical interaction detection penalty. Lastly, although we did not observe a substantial difference between equal observation weights and the observation weights proposed by \citeauthor{wan2015variable}, there are still conceptual concerns with having the observation weights depend on the fraction of missingness. As we mentioned earlier, upweighting observations with more data artificially moves the analysis in the direction of complete-case analysis, which is known to be biased under MAR and MNAR missing data mechanisms.

The positive ALS-POPs associations identified in the data example add to a growing body of literature on environmental risk factors for ALS \cite{kamel2012, mcguire1997, sutedja2008}. A major advantage of the data collected in this study is that POP concentrations were measured in plasma samples, rather than through surveys. Since ALS results from the complex interplay of multiple risks combined with neurotoxic environmental exposures (the gene-time-environment hypothesis), we contend that additional work is needed to more fully understand gene and pesticide exposure interaction \cite{paez2015amyotrophic, goutman2018emerging, al2013epidemiology}.  


\section*{Acknowledgements}
The research was supported by NSF DMS 1712933; NIH P30 CA 046591; NIEHS K23ES027221; National ALS Registry/CDC/ATSDR CDCP-DHHS-US (CDC/ATSDR 200-2013-56856); Program for Neurology Research and Discovery, University of Michigan; Robert and Katherine Jacobs Environmental Health Initiative. The authors would like to thank Alexander Rix, BS for software development support, Crystal Pacut, Jayna Duell, RN, Blake Swihart, and Daniel Burger for study support.

\section*{Data Availability Statement}
The dataset analyzed during the current study is not publicly available due to the sensitive nature of biological samples and demographic variables of the human subjects dataset, but is available from the corresponding author on reasonable request. A simulated dataset and R code supporting the conclusions of this article are available in our R package \textit{miselect}. The simulated dataset was created to closely mimic study population characteristics, especially the block correlation structure between environmental contaminants.

\section*{Conflict of Interest}
Dr. Stephen A. Goutman: scientific advisory to Biogen and ITF Pharma. DSMB service for Watermark Research Partners; additional research support from ALS Association.

\printbibliography

\end{document}


\begin{center}
\textbf{{\Large Supplementary Materials}}
\end{center}
\begin{table}[h]
\centering
\caption{\label{missingInfo}Missing data proportion for 30 variables in the ALS data. The data in total contains 266 observations with 167 cases and 99 controls.}
\begin{tabular}{p{3cm}p{3cm}p{3cm}p{3cm}}
\hline
\textbf{Variables} & \textbf{Proportion} & \textbf{Variables} & \textbf{Proportion} \\
\hline
PCB 174             & 0.508               & PCB 153             & 0.237               \\
PCB 110             & 0.429               & PCB 202             & 0.199               \\
cis-chlordane               & 0.398               & cis-nonachlor               & 0.147               \\
trans-nonachlor               & 0.380               & beta-HCH               & 0.132               \\
PCB 175             & 0.368               & PBDE 99             & 0.086               \\
trans-chlordane               & 0.365               & p,p'-DDE              & 0.083               \\
PCB 118             & 0.361               & PeCB               & 0.060               \\
PCB 180             & 0.350               & PBDE 100            & 0.056               \\
PBDE 28             & 0.305               & Education1         & 0.038               \\
PCB 138             & 0.278               & Education2         & 0.038               \\
PBDE 154            & 0.267               & PCB 151             & 0.034               \\
PBDE 153            & 0.256               & PBDE 47             & 0.015               \\
BMI                & 0.244               & Age                & 0.011               \\
BMI\_slope         & 0.244               & Sex                & 0.000               \\
PBDE 85             & 0.241               & ALS                & 0.000               \\ \hline
\end{tabular}
\end{table}

\newpage

\begin{figure}[h]
    \centering
    \includegraphics[width=1\textwidth]{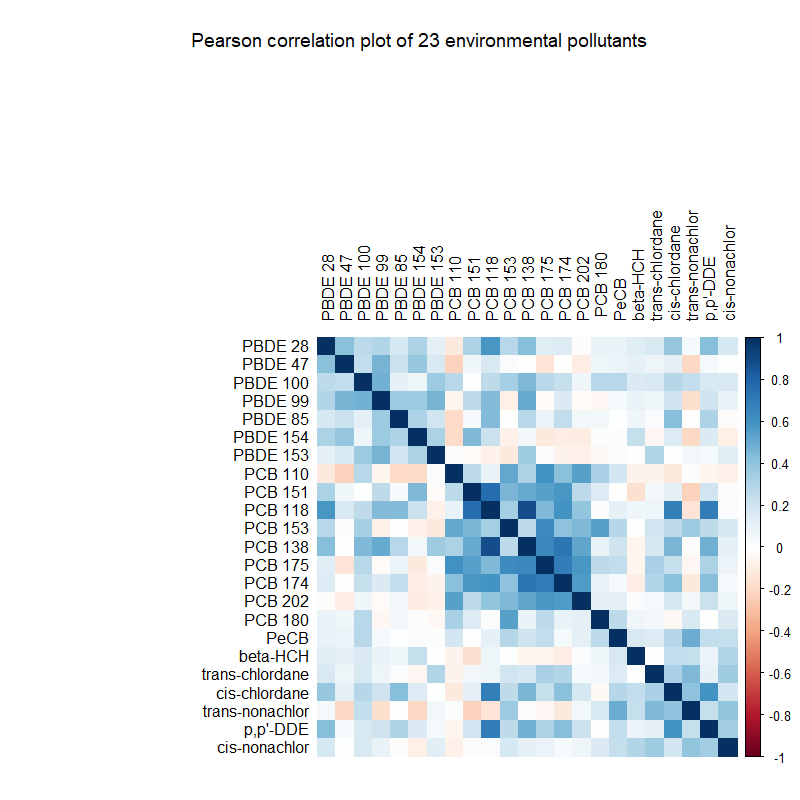}
    \caption{Pairwise Pearson correlation matrix for 23 POPs collected through the University of Michigan ALS Patient Repository. The dataset contains 266 observations (167 cases and 99 controls).}
    \label{correlation}
\end{figure}

\newpage

\begin{table}[h]
\scriptsize
\centering
\caption{\label{table:realDataAppD10} Variable selection and coefficient estimation based on 10 imputed datasets obtained from the ALS case-control study data. The data contain 167 cases and 99 controls. There are 23 environmental pollutants under consideration after adjusting for 6 demographic variables. Eleven environmental pollutants are shown in the table, and the other 12 exposures are excluded because none of the methods select them.}
\begin{tabular}{p{1.2cm}p{0.8cm}p{0.9cm}p{0.9cm}p{0.8cm}p{0.8cm}p{0.9cm}p{0.9cm}p{0.8cm}p{0.8cm}p{1cm}}
\hline
POPs & SLASSO & SaLASSO & SENET & SaENET & \begin{tabular}[c]{@{}c@{}}SLASSO\\ (w)\end{tabular} & \begin{tabular}[c]{@{}c@{}}SaLASSO\\ (w)\end{tabular} & \begin{tabular}[c]{@{}c@{}}SENET\\ (w)\end{tabular} & \begin{tabular}[c]{@{}c@{}}SaENET\\ (w)\end{tabular} & GLASSO & GaLASSO \\ \hline
PBDE 28 & - & - & 0.020 & - & - & - & 0.012 & - & - & - \\
PBDE 47 & - & - & -0.003 & - & - & - & - & - & - & - \\
PBDE 99 & - & - & -0.050 & - & - & - & -0.016 & - & - & - \\
PBDE 153 & 0.098 & - & 0.108 & - & 0.090 & 0.005 & 0.101 & - & 0.095 & - \\
PBDE 154 & - & - & -0.013 & - & - & - & - & - & - & - \\
PeCB & 0.370 & 0.695 & 0.397 & 0.642 & 0.271 & 0.596 & 0.331 & 0.592 & 0.387 & 0.672 \\
trans-chlordane & 0.081 & - & 0.151 & 0.029 & 0.064 & 0.093 & 0.128 & 0.079 & 0.115 & 0.075 \\
cis-nonachlor & 0.237 & 0.559 & 0.357 & 0.583 & 0.241 & 0.650 & 0.357 & 0.631 & 0.322 & 0.661 \\
PCB 110 & - & - & 0.007 & - & - & - & 0.004 & - & - & - \\
PCB 151 & - & 0.239 & 0.212 & 0.356 & - & 0.351 & 0.144 & 0.323 & 0.158 & 0.481 \\
PCB 202 & - & - & 0.001 & - & - & - & 0.003 & - & - & - \\
\textbf{\#selected} & \textbf{4} & \textbf{3} & \textbf{11} & \textbf{4} & \textbf{4} & \textbf{5} & \textbf{9} & \textbf{4} & \textbf{5} & \textbf{4} \\
\textbf{\#removed} & \textbf{19} & \textbf{20} & \textbf{12} & \textbf{19} & \textbf{19} & \textbf{18} & \textbf{14} & \textbf{19} & \textbf{18} & \textbf{19}\\ \hline
\end{tabular}
\end{table}